# GED: the method for group evolution discovery in social networks

Piotr Bródka · Stanisław Saganowski ·
Przemysław Kazienko

**Abstract** The continuous interest in the social network area contributes to the fast development of this field. The new possibilities of obtaining and storing data facilitate deeper analysis of the entire network, extracted social groups and single individuals as well. One of the most interesting research topic is the dynamics of social groups which means analysis of group evolution over time. Having appropriate knowledge and methods for dynamic analysis, one may attempt to predict the future of the group, and then manage it properly in order to achieve or change this predicted future according to specific needs. Such ability would be a powerful tool in the hands of human resource managers, personnel recruitment, marketing, etc. The social group evolution consists of individual events and seven types of such changes have been identified in the paper: continuing, shrinking, growing, splitting, merging, dissolving and forming. To enable the analysis of group evolution a change indicator—inclusion measure was proposed. It has been used in a new method for exploring the evolution of social groups, called Group Evolution Discovery (GED). The experimental results of its use together with the comparison to two well-known algorithms in terms of accuracy, execution time, flexibility and ease of implementation are also described in the paper.

**Keywords** Social network · Group evolution ·
Groups in social networks · Group dynamics ·
Social network analysis · Inclusion measure · GED

## 1 Introduction

Social network analysis (SNA) is gaining on importance every day, mostly because of growing number of different social networking systems and the growth of the Internet. The matter of the social networking systems may be various, starting with physical system (transportation and energy networks), through virtual systems (Internet, telecommunication, WWW), social networks, biological networks, ending on food webs and ecosystems (Barrat et al. 2008). A network analysed in this paper is a social network, which in its simplest form can be described as set of actors (network nodes) connected by relationships (network edges). Social networks have many different definitions (Hanneman and Riddle 2005, Scott 2000, Wasserman and Faust 1994, Watts and Strogatz 1998), and might have different meaning: corporate partnership networks (law partnership) (Lazega 2001), scientist collaboration networks (Newman 2001), movie actor networks, friendship network of students (Amaral et al. 2000), company director networks (Robins and Alexander 2004), sexual contact networks (Morris 1997), labour market (Montgomery et al. 1991), public health (Cattell 2001), psychology (Pagel et al. 1987), etc., but they always describe social entities and their connections.

P. Bródka (✉) · S. Saganowski · P. Kazienko
Institute of Informatics, Wrocław University of Technology,
Wyb.Wyspiańskiego 27, 50-370 Wrocław, Poland
e-mail: piotr.brodka@pwr.wroc.pl

S. Saganowski
e-mail: stanislaw.saganowski@pwr.wroc.pl

P. Kazienko
e-mail: kazienko@pwr.wroc.pl

P. Bródka · P. Kazienko
Research Engineering Center Sp. z o.o.,
ul. Strzegomska 46B, 53-611 Wrocław, Poland

Group extraction and their evolution are among the topics which arouse the greatest interest in the domain of social network analysis. However, while the group extraction methods for social networks are being developed very dynamically, the methods of group evolution discovery and analysis are still 'uncharted territory' on the social network analysis map. Therefore, the new method for the group evolution discovery called GED is proposed in this paper. Additionally, the results of comparison with two other methods of group evolution discovery are presented. It should also be mentioned that this article is a continuation and significant extension of research presented in (Bródka et al. 2011a, b).

The article is organized as follows: Sect. 2 describes the related work; Sect. 3 presents some basic concepts and definitions like temporal social network, group or social position measure, which help to understand the environment and the new method itself. In Sect. 4 group evolution and its component steps are described, together with the new method of group evolution extraction based on member position in the social network called GED. The results of experimental studies are presented in Sect. 5 and concluded in Sect. 6.

## 2 Related work

The most available social networks for investigation are online social networks (Garton et al. 1997), web-based social networks (Golbeck and Hendler 2006), computer-supported social networks (Wellman et al. 1996) or virtual social networks. The reason for this is the simple and continuous way to collect communication or collaboration data, from which we can extract these social networks. The source data can be found in various systems, e.g. bibliographic data (Girvan and Newman 2002), blogs (Agarwal and Galan 2010), photos sharing systems like Flickr (Kazienko et al. 2011), email systems (Tyler et al. 2003), telecommunication data (Blondel et al. 2008, Kazienko et al. 2009a, b), social services like Twitter (Huberman et al. 2009) or Facebook (Ellison et al. 2007; Traud et al. 2009), video sharing systems like YouTube (Cheng et al. 2008) and many more. Obtaining the data from these sources for a longer period allows to explore more than one social network in specific snapshots of time. Using proper techniques, it is possible to evaluate changes occurring in the social network over time. The most interesting are analysis on the following changes of social groups (communities) extracted from the social network. This enables to describe and observe the whole dynamic process of group evolution.

In recent years, several methods for tracking changes in social groups have been proposed. Sun et al. (2007) have introduced GraphScope, Chakrabarti et al. (2006) have presented another original approach, Lin et al. (2008) have provided the framework called FacetNet, Kim and Han in (Kim 2009) have introduced the concept of nano-communities, Hopcroft et al. (Hopcroft 2004) have also investigated group evolution, however no method which can be implemented have been provided. Two methods evaluated in this article are described below in greater detail.

Asur et al. 2007 have proposed a simple approach for investigating group evolution over time. At first, groups are extracted in each time frame, then comparing the size and overlapping of every possible pair of groups in consecutive time steps, the events involving those groups are assigned. When none of the nodes in the group from time step $T_i$ occurs in the following time frame $T_{i+1}$, Asur et al. have described this situation as *dissolve* of the group. In opposite to dissolve, if none of the nodes in the group from time frame $T_i$ was present in the previous time frame $T_{i-1}$, a group is marked as *new born*. The group *continues* its existence when identical occurrence of the group in the consecutive time frames is found. Case, when two groups from time step $T_{i-1}$ joined together overlap or overlap each other with more than a given percentage of the single group in time frame $T_i$, is called *merge*. In the opposite case, when two groups from time frame $T_i$ joined together overlap greater than a given part of the single group in time frame $T_{i+1}$, the event is marked as *split*. Asur et al. did not specify what method has been used for group extraction or if the method works for overlapping groups.

Palla et al. (2007) have used clique percolation method (CPM) (Palla et al. 2005; Derényi et al. 2005), which allows groups to overlap. Thanks to this feature analysing changes in groups over time is very simple. Networks at two consecutive time frames $T_i$ and $T_{i+1}$ are merged into a single graph $Q(T_i, T_{i+1})$ and groups are extracted using the CPM method. Next, the communities from time frames $T_i$ and $T_{i+1}$, which are the part of the same group from the joined graph $Q(T_i, T_{i+1})$, are considered to be matching. It may happen that more than two communities are contained in the same group. Then, matching is performed based on the value of their relative overlap sorted in descending order. Possible events between groups are: growth, contraction, merging, splitting, birth and death. Using the CPM method allowed Palla et al. to investigate evolution in overlapping groups, which can be extracted from the directed as well as weighted network.

## 3 Group evolution discovery

Before the Group Evolution Method (GED) is presented, it is necessary to describe a few general concepts related to social networks: a temporal social network, social group, group evolution, and social position—a centrality measure.

### 3.1 Temporal social network

A temporal social network TSN is a list of following time frames (time windows) $T$. Each time frame is in fact a single social network SN($V,E$), where $V$ is a set of vertices and $E$ is a set of directed edges $\langle x,y \rangle : x,y \in V$

$$\begin{aligned} &TSN = \langle T_1, T_2, \ldots, T_m \rangle, \quad m \in N \\ &T_i = SN_i(V_i, E_i), \quad i = 1, 2, \ldots, m \\ &E_i = \langle x, y \rangle : x, y \in V_i, \quad i = 1, 2, \ldots, m \end{aligned} \quad (1)$$

An example of a temporal social network TSN is presented in Fig. 1. It consists of five time frames, and each time frame is a separate social network created from data gathered in the particular interval of time. In the simplest case, one interval starts when the previous interval ends, but based on author's needs the intervals may overlap by a set of time or can even contain full history of previous time frames in the aggregated form.

### 3.2 Social group

There is no universally acceptable definition of groups in the social network (Coleman 1964; Fortunato 2010). Nevertheless, there are several of them, which are used depending on the authors' needs (Coleman 1964; Freeman 2004; Kottak 2004; Fortunato 2010). In addition, some of them cannot even be called definitions because they introduce only some criteria for the group existence. In the biological terminology, a group, often also called a community is a collection of cooperating organisms, sharing a common environment. In sociology, in turn, it is traditionally defined as a group of people living and cooperating in a single location. However, due to the fast growing and spreading Internet, the concept of social community has lost its geographical limitations. Overall, a general idea of the social community is a group in a given population, whose members more frequently collaborate with each other rather than with other members of this population (the entire social network). The concept of the group (social community) can be easily transposed to the graph theory, in which the social network is a graph and a group is a subset of vertices with high density of edges inside the group, and lower edge density between nodes from two separate groups. However, another problem arises in the quantitative definition of a community. Anyway, most definitions of groups are built based on the general idea presented above. Additionally, groups can also be algorithmically determined, as the outcome of the specific clustering algorithm, i.e. without a precise a priori definition (Moody and White 2003). In this paper, the following definition will be used: a group $G$ extracted from the social network SN($V,E$) is a subset of vertices from $V(G \subseteq V)$, extracted using any community extraction method (clustering algorithm).

### 3.3 Group evolution

Group evolution is a sequence of events (changes) succeeding each other in the consecutive time windows (time frames) within the social network. Palla et al. (2007) and Asur et al. (2007) have proposed some types of events but their lists were incomplete. Thus, in this paper, the possible list of events in social group evolution was extended. Seven independent types of events have been identified changing the state of a group or groups between two following time windows (see Fig. 2):

1. *Continuing* (*stagnation*) A group continues its existence, when two groups in the consecutive time windows are identical or when two groups differ only by few nodes but their size remains the same.

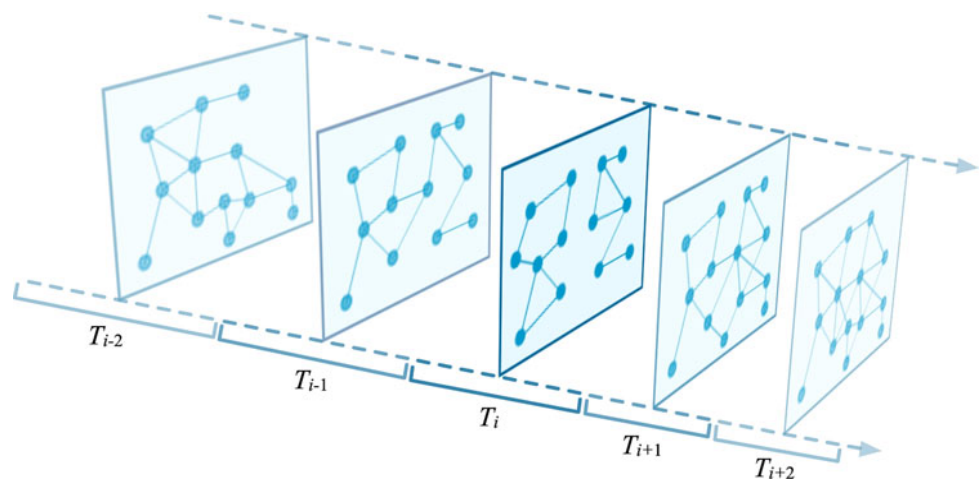

**Fig. 1** The example of temporal social network consisting of five time frames

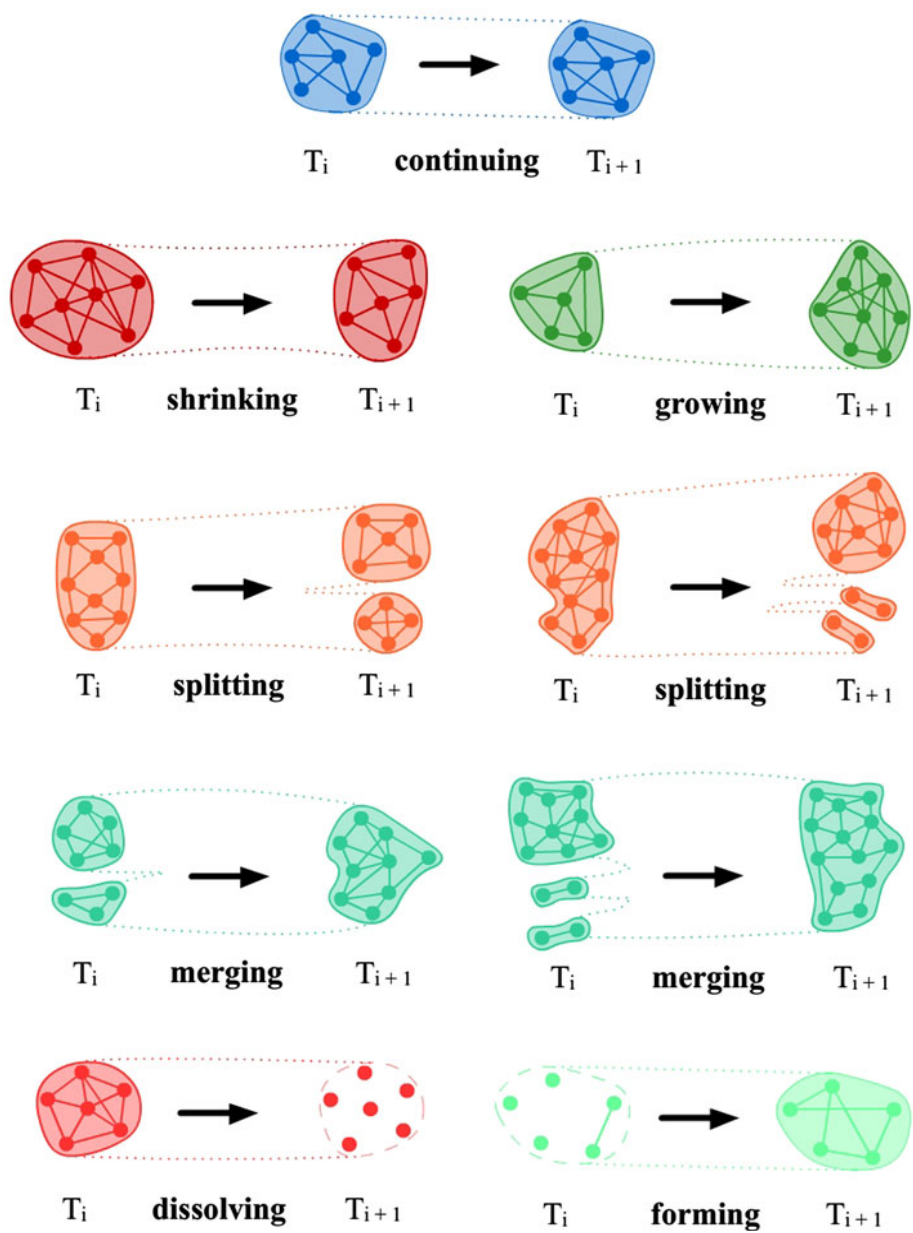

Fig. 2 Seven possible types of events in the group evolution

2. *Shrinking* A group shrinks when some nodes have left the group, making its size smaller than in the previous time window. A group can shrink slightly, i.e. by a few nodes or greatly losing most of its members.
3. *Growing* (*opposite to shrinking*) A group grows when some new nodes have joined the group, making its size bigger than in the previous time window. A group can grow slightly as well as significantly, doubling or even tripling its size.
4. *Splitting* A group splits into two or more groups in the next time window $T_{i+1}$, when some groups from time frame $T_{i+1}$ consist of members of one group from the previous time frame $T_i$. We can distinguish two types of splitting: (1) *equal split*, which means the contribution of all resulting groups in the splitting group is almost the same and (2) *unequal split* when one of the final groups has much greater contribution in the splitting group, which in turn for this greater group might be similar to shrinking.
5. *Merging* (*reverse to splitting*) A group has been created by merging several other groups when one group from time frame $T_{i+1}$ consists of two or more groups from the previous time frame $T_i$. The merge, just like the split, might be (1) *equal*, when the contribution of all source groups in the merged, target group is almost the same, or (2) *unequal*, if one of the groups has much greater contribution into the merged group. In second case, for the biggest group the merging might be similar to growing.

6. *Dissolving* happens when a group ends its life and does not occur in the next time window at all, i.e. its members have vanished or stopped communicating with each other and are scattered among the rest of the groups.

7. *Forming* of the new group (opposite to dissolving) occurs when a group, which did not exist in the previous time window $T_i$, appears in next time window $T_{i+1}$. When a group remains inactive over several time frames, such case is treated as dissolving of the first group and forming again of the second, new one.

### 3.4 Social position

To discover group evolution, the GED method described below takes into account both, the quantity and quality of the group members. To express group member quality, one of the centrality measures may be used. In the experimental studies, social position SP measure (Bródka et al. 2009; Kazienko et al. 2009a, b) was utilized.

The social position for the network SN(V,E) is calculated in the iterative way, as follows:

$$\mathrm{SP}_{n+1}(x) = (1-\varepsilon) + \varepsilon \cdot \sum_{y \in V} \mathrm{SP}_n(y) \cdot C(y \to x), \qquad (2)$$

where $\mathrm{SP}_{n+1}(x)$ and $\mathrm{SP}_n(x)$ is the social position of member $x$ after the $n+1$st and $n$th iteration, respectively, and $\mathrm{SP}_0(x) = 1$ for each $x \in V$; $\varepsilon$ is the fixed coefficient from the range (0;1); $C(y \to x)$ is the commitment function, which expresses the strength of the relation from $y$ to $x$—the weight of edge $\langle y, x \rangle$.

Social position can also be calculated only for a part of the entire social network, i.e. only within a certain social group $G$. Then the sum in (2) is over $y \in G$. For detailed information about social position measure, how to calculate and implement it see Bródka et al. (2009), Kazienko et al. (2009a, b), Musial et al. (2009).

## 4 GED: a method for group evolution discovery in the social network

To discover group evolution in the social network a new method called GED (Group Evolution Discovery) was developed. The most important component of this method is a new measure called *inclusion*. This measure allows to evaluate the inclusion of one group in another. Therefore, inclusion $I(G_1, G_2)$ of group $G_1$ in group $G_2$ is calculated as follows:

$$I(G_1, G_2) = \overbrace{\frac{|G_1 \cap G_2|}{|G_1|}}^{\text{group quantity}} \cdot \underbrace{\frac{\sum_{x \in (G_1 \cap G_2)} \mathrm{SP}_{G_1}(x)}{\sum_{x \in (G_1)} \mathrm{SP}_{G_1}(x)}}_{\text{group quality}} \qquad (3)$$

where $\mathrm{SP}_{G_1}(x)$ is the value of social position of node $x$ in group $G_1$.

Of course, instead of social position any other measure which indicates member position within the community can be used, e.g. centrality degree, betweenness degree, page rank, etc. The second factor in (3) would have to be adapted accordingly in such case. However, after analysing the complexity of computation and diversity of measure values (Musial et al. 2009), authors have decided to utilize social position measure as an example in this paper.

As it was mentioned before, the GED method, used to discover group evolution, respects both the quantity and quality of the group members. The *quantity* is reflected by the first part of the *inclusion* measure, i.e. what portion of members from group $G_1$ is in group $G_2$, whereas the *quality* is expressed by the second part of the *inclusion* measure, namely what contribution of important members from group $G_1$ is in $G_2$. It provides a balance between the groups that contain many of the less important members and groups with only few but key members.

One might say that inclusion measure is "unfair" for not identical groups, because if the community differs even by only one member, inclusion is reduced through not having all nodes and also through not having social position of those nodes. Indeed, it is slightly "unfair" (or rather strict), but using member position within the community calculated on the basis of users relations makes *inclusion* to focus not only on nodes (members) but also on edges (relations) giving great advantage over methods which are using only members' overlapping for event identification (group quantity factor in inclusion measure).

It is assumed that only one event may occur for two groups ($G_1$, $G_2$) in the consecutive time frames; however, one group in time frame $T_i$ may be involved in several events with different groups in $T_{i+1}$.

The procedure for the Group Evolution Method (GED) is as follows:

GED: Group Evolution Discovery method

**Input**: Temporal social network *TSN*, in which groups are extracted by any community detection algorithm separately for each time frame $T_i$ and any user importance measure is calculated for each group.

1. For each pair of groups $<G_1, G_2>$ in consecutive time frames $T_i$ and $T_{i+1}$ inclusion $I(G_1, G_2)$ for $G_1$ in $G_2$ and $I(G_2, G_1)$ for $G_2$ in $G_1$ is computed according to (3).
2. Based on both inclusions $I(G_1, G_2)$, $I(G_2, G_1)$ and sizes of both groups only one type of event may be identified:

   (a) *Continuing* $I(G_1, G_2) \geq \alpha$ and $I(G_2, G_1) \geq \beta$ and $|G_1| = |G_2|$

   (b) *Shrinking* $I(G_1, G_2) \geq \alpha$ and $I(G_2, G_1) \geq \beta$ and $|G_1| > |G_2|$ OR $I(G_1, G_2) < \alpha$ and $I(G_2, G_1) \geq \beta$ and $|G_1| \geq |G_2|$ and there is only one match (matching event) between $G_2$ and all groups in the previous time window $T_i$

   (c) *Growing* $I(G_1, G_2) \geq \alpha$ and $I(G_2, G_1) \geq \beta$ and $|G_1| < |G_2|$ OR $I(G_1, G_2) \geq \alpha$ and $I(G_2, G_1) < \beta$ and $|G_1| \leq |G_2|$ and there is only one match (matching event) between $G_1$ and all groups in the next time window $T_{i+1}$

   (d) *Splitting* $I(G_1, G_2) < \alpha$ and $I(G_2, G_1) \geq \beta$ and $|G_1| \geq |G_2|$ and there is more than one match (matching event) between $G_2$ and all groups in the previous time window $T_i$

   (e) *Merging* $I(G_1, G_2) \geq \alpha$ and $I(G_2, G_1) < \beta$ and $|G_1| \leq |G_2|$ and there is more than one match (matching event) between $G_1$ and all groups in the next time window $T_{i+1}$

   (f) *Dissolving* for $G_1$ in $T_i$ and each group $G_2$ in $T_{i+1}$ $I(G_1, G_2) < 10\%$ and $I(G_2, G_1) < 10\%$

   (g) *Forming* for $G_2$ in $T_{i+1}$ and each group $G_1$ in $T_i$ $I(G_1, G_2) < 10\%$ and $I(G_2, G_1) < 10\%$

The scheme, which facilitates understanding of the event selection (identification) for the pair of groups in the GED method is presented in Fig. 3.

$\alpha$ and $\beta$ are the GED method parameters, which can be used to adjust the method to the particular social network and community detection method. According to experimental analysis (see Sect. 5) the authors suggest the values of $\alpha$ and $\beta$ to be from the range [50%;100%]

Based on the list of extracted events, which have occurred for the selected group between each two successive time frames, the whole group evolution process may be created.

In the sample social network in Fig. 4 and Table 1, its lifetime consists of eight time windows. The group forms in $T_2$, then it grows in $T_3$ by gaining some new nodes, next it splits into two groups in $T_4$, afterwards the bigger group is shrinking in $T_5$ by losing one node, both groups continue over $T_6$ and they both merge with the third group in $T_7$, finally the group dissolves in $T_8$.

## 5 Experiments

The experiments were conducted on the data gathered from Wroclaw University of Technology email communication. The whole data set was collected within period of February 2006–October 2007 and consists of 5,845 nodes (university distinct email addresses) and 149,344 edges (emails send from one address to another).

The temporal social network consisted of fourteen 90-day time frame extracted from this source data. Time frames have the 45-day overlap, i.e., the first time frame begins on the 1st day and ends on the 90th day, the second begins on the 46th day and ends on the 135th day and so on.

### 5.1 Experiment based on overlapping groups extracted by CPM

In the first experiment, as a method for group extraction, CPM was utilized (http://www.cfinder.org/). The groups were discovered for $k = 6$ and for the directed and unweighted social network. The CPM algorithm has extracted from 80 to 136 groups for different time windows (avg. 112 per time window). The average size of the group was 19 nodes. The smallest group had size of 6, because of the $k$ parameter and the biggest one was of 613 in time window 10.

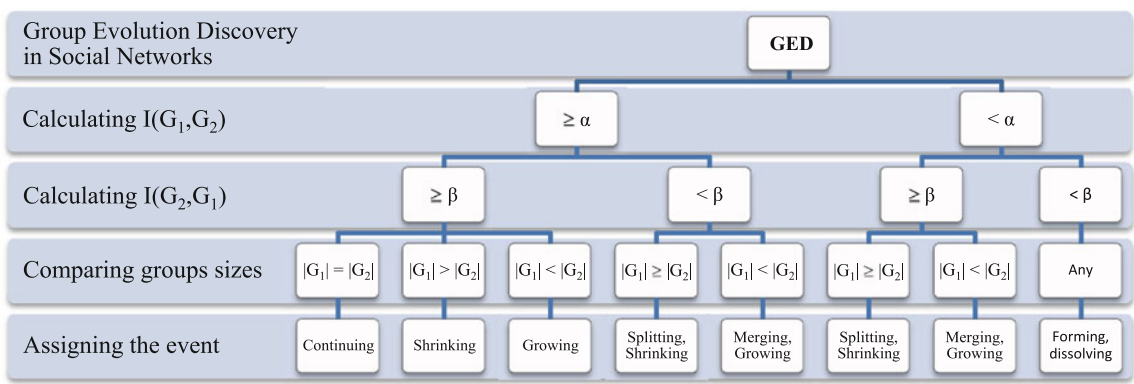

**Fig. 3** The decision tree for assigning the event type to a pair of groups

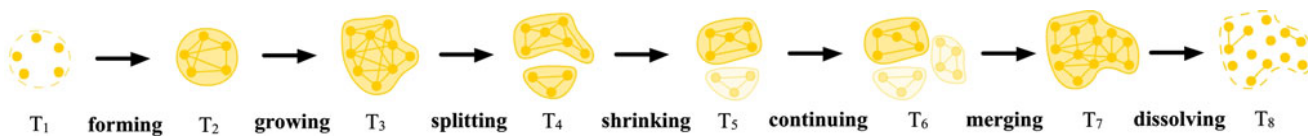

Fig. 4 Changes over time for the single group

Table 1 Changes over time for the single group

| Event type | Group in $T_2$ | Event type | Group in $T_3$ | Event type | Group in $T_4$ | Event type | Group in $T_5$ | Event type | Group in $T_6$ | Event type | Group in $T_7$ | Event type |
|---|---|---|---|---|---|---|---|---|---|---|---|---|
| Form | $G_1$ | Growth | $G_1$ | Split | $G_2$ | Shrink | $G_2$ | Continue | $G_2$ | Merge | $G_5$ | Dissolve |
| - | - | - | | | $G_3$ | Continue | $G_3$ | Continue | $G_3$ | | | |
| - | - | - | - | - | - | - | - | Form | $G_4$ | | | |

### 5.1.1 Method by Asur et al.

This method has been implemented in T-SQL language. The authors have suggested to set 30 or 50% as an overlapping threshold for merge and split. In the experiment, the threshold was set to 50%. It took more than 5.5 h to calculate events between groups in all 14 time frames. The total number of events found by Asur et al. method was 1,526, out of which 90 were continuation, 18—forming, 29—dissolving, 703—merging and 686 were splitting.

Such a small number of continuing events is caused by the very rigorous condition, which requires the groups to remain unchanged. Small amount of forming (dissolving) events came from another strong condition, which states that none of the nodes from the considered group can exist in the network in previous (following) time window. A huge number of merging (splitting) events is a result of low overlapping threshold for merge (split).

However, it has to be noticed that these numbers are slightly overestimated. The method by Asur et al. allows one pair of groups to assign more than one type of events. This leads to anomalies when, e.g. the group no. 1 in time window no. 1 ($T_1$) is continuing in group no. 2 in $T_2$ and simultaneously also merging with group no. 13 from $T_1$ into group no. 2 in $T_2$. This should not happen if the condition for continuing is so rigorous.

The total number of anomalies is 128 cases, 8% of all results. More than a half of these cases are groups with *split* and *merge* event into another group at the same time. The rest of the cases are even worse, because one group has *continue* and *split* or *merge* event into another group simultaneously. Therefore, the total number of "distinct" events found by Asur et al. was as many as 1,398.

All these unexpected cases revealed a significant weakness of the method by Asur et al.

### 5.1.2 Method by Palla et al.

The method by Palla et al. has been implemented in T-SQL, but it required much more preparations. Apart from extracting groups in all time windows, yet another group extraction was needed. The data from two consecutive time windows were merged into a single graph, from which groups were extracted by means of the CPM method. As easy to count, the group extraction had to be performed additional 13 times; some of them took only 5 min to calculate, but there were also some lasting up to 2 days.

Palla et al. have designed their method in order to find all matching pairs of groups, even if they overlap in the slightest way, sharing only one node. The great advantage of this approach is that no event will be ignored. However, if one takes into account the fact that Palla et al. only showed which event types may occur (and did not provide the algorithm to assign them), analysis of the group evolution during its life is very difficult and cumbersome. Each case of assigning event must be considered individually over a huge number of possibilities. As a result, it is very hard to find the key match. Moreover, Palla et al. did not explain how to choose the best match for the analysed groups or how to assign the event type. The authors only defined the case when there is the single highest overlapping for each group.

The total number of matched pairs found by Palla's et al. method was 9,797, out of which 4,183 pairs (42.7%) had an overlap >0%. The authors did not specify how to interpret the rest of the group that matched with overlap of 0%, but intuition suggests to omit these cases. There were 90 cases when matched pairs had overlap equal 100%, which corresponds to *continuation* event in the Asur et al. method.

**Table 2** The results of the GED computation on overlapping groups extracted by CPM

| Threshold | | Number of events | | | | | | | Total |
|---|---|---|---|---|---|---|---|---|---|
| α | β | Form | Dissolve | Shrink | Growth | Continue | Split | Merge | |
| 50 | 50 | 122 | 186 | 204 | 180 | 127 | 517 | 398 | 1,734 |
| 50 | 60 | 122 | 186 | 204 | 173 | 124 | 464 | 405 | 1,678 |
| 50 | 70 | 122 | 186 | 202 | 157 | 124 | 400 | 421 | 1,612 |
| 50 | 80 | 122 | 186 | 203 | 149 | 122 | 311 | 429 | 1,522 |
| 50 | 90 | 122 | 186 | 199 | 154 | 122 | 279 | 424 | 1,486 |
| 50 | 100 | 122 | 186 | 199 | 156 | 122 | 261 | 422 | 1,468 |
| 60 | 50 | 122 | 186 | 190 | 177 | 124 | 531 | 359 | 1,689 |
| 60 | 60 | 122 | 186 | 191 | 170 | 120 | 475 | 366 | 1,630 |
| 60 | 70 | 122 | 186 | 187 | 152 | 119 | 409 | 384 | 1,559 |
| 60 | 80 | 122 | 186 | 187 | 144 | 117 | 314 | 392 | 1,462 |
| 60 | 90 | 122 | 186 | 181 | 148 | 117 | 277 | 388 | 1,419 |
| 60 | 100 | 122 | 186 | 179 | 149 | 117 | 259 | 387 | 1,399 |
| 70 | 50 | 122 | 186 | 179 | 176 | 123 | 543 | 284 | 1,613 |
| 70 | 60 | 122 | 186 | 180 | 170 | 119 | 486 | 286 | 1,549 |
| 70 | 70 | 122 | 186 | 177 | 156 | 113 | 418 | 298 | 1,470 |
| 70 | 80 | 122 | 186 | 174 | 149 | 111 | 317 | 305 | 1,364 |
| 70 | 90 | 122 | 186 | 165 | 150 | 111 | 277 | 304 | 1,315 |
| 70 | 100 | 122 | 186 | 161 | 152 | 111 | 259 | 302 | 1,293 |
| 80 | 50 | 122 | 186 | 172 | 169 | 120 | 553 | 233 | 1,555 |
| 80 | 60 | 122 | 186 | 173 | 154 | 117 | 495 | 235 | 1,482 |
| 80 | 70 | 122 | 186 | 170 | 137 | 111 | 426 | 244 | 1,396 |
| 80 | 80 | 122 | 186 | 165 | 127 | 97 | 324 | 251 | 1,272 |
| 80 | 90 | 122 | 186 | 157 | 128 | 96 | 276 | 250 | 1,215 |
| 80 | 100 | 122 | 186 | 152 | 129 | 96 | 257 | 249 | 1,191 |
| 90 | 50 | 122 | 186 | 172 | 169 | 120 | 553 | 199 | 1,521 |
| 90 | 60 | 122 | 186 | 174 | 152 | 117 | 494 | 198 | 1,443 |
| 90 | 70 | 122 | 186 | 171 | 132 | 111 | 425 | 199 | 1,346 |
| 90 | 80 | 122 | 186 | 165 | 121 | 96 | 324 | 203 | 1,217 |
| 90 | 90 | 122 | 186 | 154 | 123 | 91 | 276 | 199 | 1,151 |
| 90 | 100 | 122 | 186 | 148 | 123 | 91 | 257 | 199 | 1,126 |
| 100 | 50 | 122 | 186 | 176 | 167 | 120 | 549 | 185 | 1,505 |
| 100 | 60 | 122 | 186 | 177 | 149 | 117 | 491 | 183 | 1,425 |
| 100 | 70 | 122 | 186 | 173 | 127 | 111 | 423 | 180 | 1,322 |
| 100 | 80 | 122 | 186 | 166 | 116 | 96 | 323 | 179 | 1,188 |
| 100 | 90 | 122 | 186 | 154 | 117 | 91 | 276 | 173 | 1,119 |
| 100 | 100 | 122 | 186 | 148 | 115 | 90 | 257 | 173 | 1,091 |

### 5.1.3 The GED method

The GED method has also been implemented in T-SQL language. The method has been run frequently with different values of α and β thresholds to analyse the influence of these parameters on the method (see Table 2). The time needed for a single run was about 6 min. The lowest checked value for the thresholds was set to 50%, which guarantees that at least a half of the considered groups are included in the matched group. The highest possible value is of course 100% which means the studied group is identical to the matched group. The thresholds for the *forming* and *dissolving* event were set to 10% based on average group size and intuition.

While analysing Table 2 and Figs. 5, 6, 7, 8, 9, 10, it can be observed that with the increase of α and β thresholds, the total number of events is decreasing: when α and β equal 50% this number is 1,734, and with thresholds equal 100% the number is only 1,091. It means that the parameters α and β can be used to filter results, preserving from events where groups are highly overlapped. Another advantage of having parameters is the possibility to adjust the results to one's needs. The linear increase of threshold α causes close to linear reduction in the number of merging events. In contrast, with linear increase of threshold β, the number of splitting events decreases in almost linear way. As a consequence of the algorithm structure, raising the thresholds makes it difficult to match the groups (see Fig. 3). Furthermore, dissolving events occur more frequently than forming events. The main reason is the fact that the last time window covers only the period of summer holidays, and as a result the email exchange is very low for that time. This causes the groups to be small and have low density.

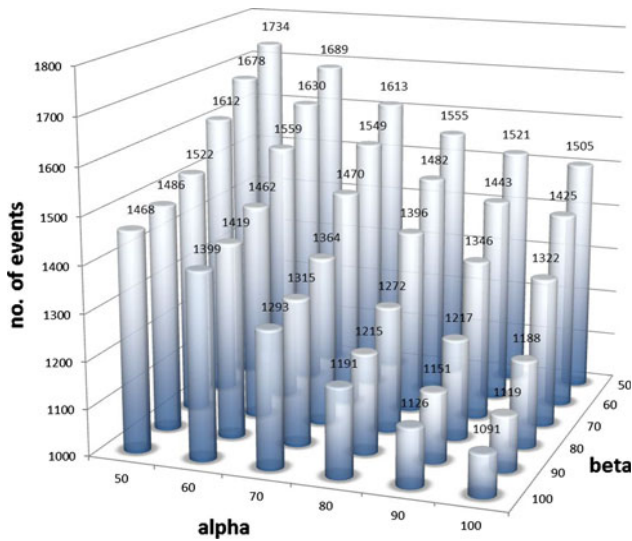

Fig. 5 Alpha and beta influence on the number of events

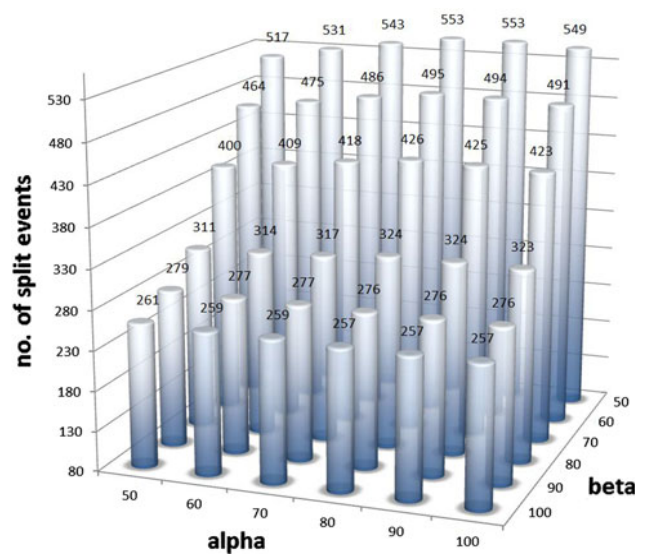

Fig. 7 Alpha and beta influence on the number of split events

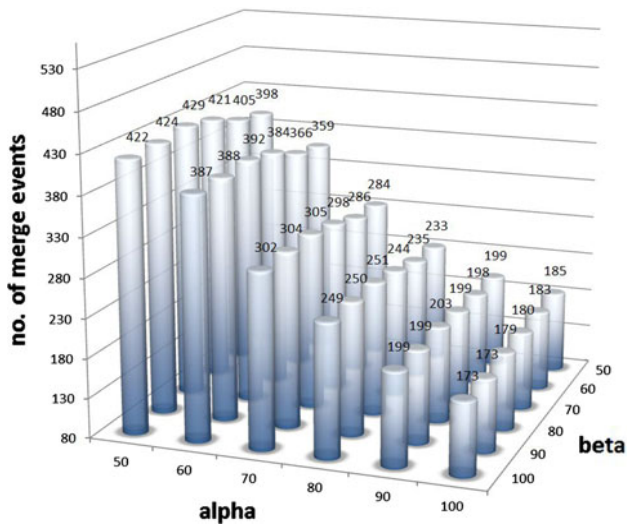

Fig. 6 Alpha and beta influence on the number of merge events

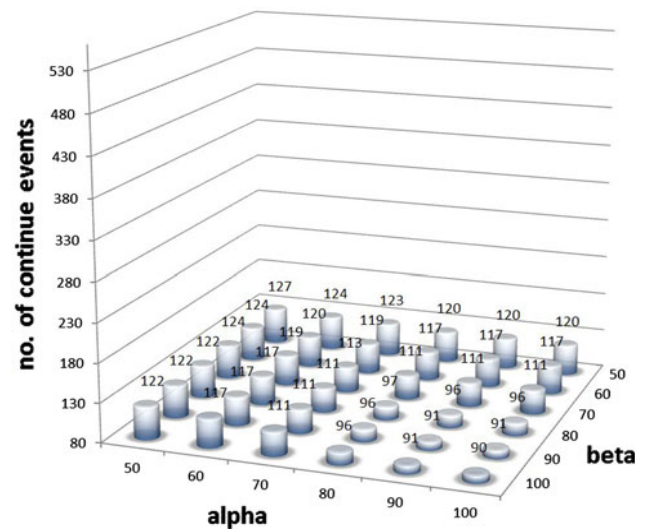

Fig. 8 Alpha and beta influence on the number of continue events

Overall, the GED method found 90 continue events when both inclusions of groups (α and β) are equal to 100%.

### 5.1.4 Differences between the GED method and the method by Asur et al.

As already mentioned, the computation time for Asur et al. method was more than 5.5 h, while for GED it took less than 4 h to calculate the whole Table 2. The single run of the GED method lasted less than 6 min, so it is over 50 times faster than the method by Asur et al.

The GED method with thresholds equal to 50% has found 721 events which the method by Asur et al. has not discovered at all. Such a big lacuna in results obtained with

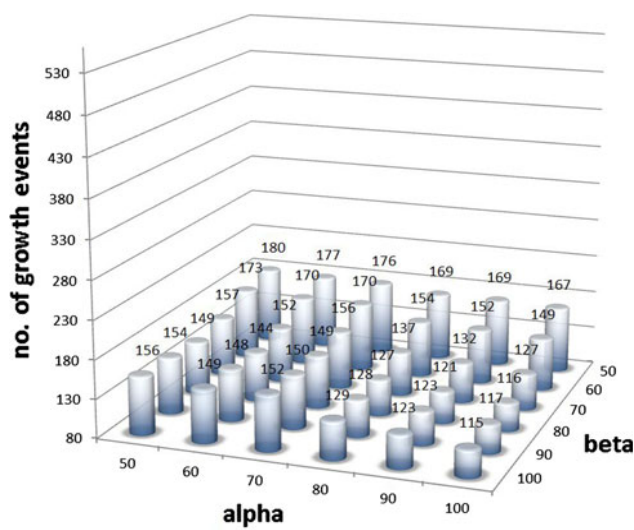

**Fig. 9** Alpha and beta influence on the number of growth events

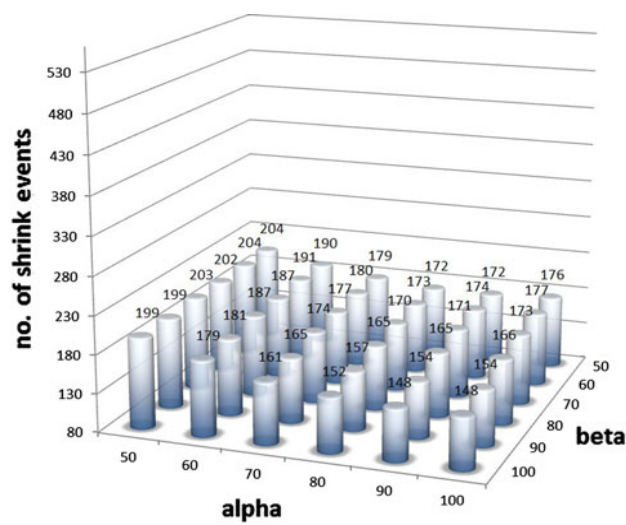

**Fig. 10** Alpha and beta influence on the number of shrink events

Asur et al. method is caused mostly by its rigorous conditions for assigning events and almost no flexibility of the method. On the other hand, Asur et al. method has found 399 events which the GED method with thresholds 50% has not. However, it is not treated as a defect in GED's results because all these events had both inclusions below 50%, therefore, the GED algorithm skipped them on purpose (because of thresholds' values). To prove this, the GED method was run with thresholds equal to 10% and this time none of events found by Asur et al. method were omitted by the GED method.

Furthermore, Asur et al. did not introduce the shrinking and growing events, which affects assigning splitting and merging events or, in the worst case, missing the event. If two groups in the successive time windows differ only by one node, they will not be treated as continuation (since the overlapping is below 100%) and it might not be treated as merging (splitting), if there is no other group fulfilling the requirements for merging (splitting). Such a case is not possible in GED, which through the change of inclusion thresholds allows to adjust the results to user's needs.

The above analysis proves that the GED method is not only faster but also more accurate and much more flexible than method by Asur et al.

### 5.1.5 Differences between the GED method and the method by Palla et al.

As noted before, the method by Palla et al. needs additional preparations to run the experiment, which lasted almost a week; therefore, the GED method is incomparable faster, despite additional calculations of user importance measures required.

The great advantage of the method by Palla et al. is catching all matching pairs of groups. As in the case of comparing the GED method with the algorithm by Asur et al., Palla et al. method found more matched pairs than the GED method with thresholds at the level of 50%. Again, it is not treated as a defect in GED's results since all these events had both inclusions below 50%. To confirm that, the results obtained with GED on thresholds equal 10% have been compared, and this time all matched pairs found by Palla et al. and not found by the GED method had inclusions below 10%. What is more, the GED method found 308 events (forming and dissolving) missed by Palla et al. method.

Another problem with Palla et al. method is the lack of the algorithm for assigning events. It is very difficult and time consuming to identify an event for the group in the next time window, not to mention for all 14 time windows. So, the GED method with its automatic event assignment is much more useful and convenient.

Summing up, the GED method is not comparable when it comes to execution time. It is also definitely more specific in assigning events and therefore much more effective and accurate for tracking group evolution. The method by Palla et al. was helpful only in checking if the GED method found all events between the groups.

### 5.2 Experiment based on disjoint groups extracted by *Blondel*

For the second experiment the method called Fast Modularity Optimization or *Blondel* was used (Blondel et al. 2008).

#### 5.2.1 Method by Asur et al.

The method provided by Asur et al. needed almost 6 h to calculate events between groups for all 14 time windows.

GED: the method for group evolution discovery in social networks

**Table 3** The results of the GED identification process for disjoint groups extracted by Blondel

| Threshold | | Number of events | | | | | | | Total |
|---|---|---|---|---|---|---|---|---|---|
| α | β | Form | Dissolve | Shrink | Growth | Continue | Split | Merge | |
| 50 | 50 | 39 | 23 | 187 | 167 | 135 | 411 | 269 | 1,231 |
| 50 | 60 | 39 | 23 | 181 | 161 | 135 | 378 | 275 | 1,192 |
| 50 | 70 | 39 | 23 | 179 | 156 | 135 | 338 | 280 | 1,150 |
| 50 | 80 | 39 | 23 | 178 | 153 | 135 | 294 | 283 | 1,105 |
| 50 | 90 | 39 | 23 | 164 | 143 | 134 | 250 | 293 | 1,046 |
| 50 | 100 | 39 | 23 | 154 | 143 | 134 | 224 | 293 | 1,010 |
| 60 | 50 | 39 | 23 | 181 | 166 | 135 | 417 | 237 | 1,198 |
| 60 | 60 | 39 | 23 | 176 | 159 | 134 | 383 | 244 | 1,158 |
| 60 | 70 | 39 | 23 | 174 | 155 | 134 | 338 | 247 | 1,110 |
| 60 | 80 | 39 | 23 | 171 | 151 | 134 | 294 | 251 | 1,063 |
| 60 | 90 | 39 | 23 | 156 | 140 | 133 | 250 | 262 | 1,003 |
| 60 | 100 | 39 | 23 | 148 | 140 | 133 | 218 | 262 | 963 |
| 70 | 50 | 39 | 23 | 169 | 164 | 134 | 429 | 216 | 1,174 |
| 70 | 60 | 39 | 23 | 163 | 158 | 131 | 396 | 219 | 1,129 |
| 70 | 70 | 39 | 23 | 164 | 154 | 130 | 345 | 221 | 1,076 |
| 70 | 80 | 39 | 23 | 159 | 150 | 130 | 299 | 225 | 1,025 |
| 70 | 90 | 39 | 23 | 144 | 139 | 129 | 245 | 236 | 955 |
| 70 | 100 | 39 | 23 | 137 | 138 | 129 | 204 | 237 | 907 |
| 80 | 50 | 39 | 23 | 162 | 165 | 134 | 436 | 180 | 1,139 |
| 80 | 60 | 39 | 23 | 157 | 158 | 130 | 402 | 178 | 1,087 |
| 80 | 70 | 39 | 23 | 156 | 152 | 129 | 350 | 176 | 1,025 |
| 80 | 80 | 39 | 23 | 151 | 147 | 127 | 304 | 177 | 968 |
| 80 | 90 | 39 | 23 | 138 | 140 | 126 | 235 | 184 | 885 |
| 80 | 100 | 39 | 23 | 128 | 140 | 126 | 191 | 184 | 831 |
| 90 | 50 | 39 | 23 | 157 | 172 | 133 | 442 | 126 | 1,092 |
| 90 | 60 | 39 | 23 | 153 | 161 | 129 | 407 | 124 | 1,036 |
| 90 | 70 | 39 | 23 | 152 | 152 | 128 | 355 | 118 | 967 |
| 90 | 80 | 39 | 23 | 146 | 139 | 126 | 310 | 116 | 899 |
| 90 | 90 | 39 | 23 | 133 | 130 | 121 | 228 | 114 | 788 |
| 90 | 100 | 39 | 23 | 116 | 131 | 121 | 178 | 113 | 721 |
| 100 | 50 | 39 | 23 | 160 | 168 | 133 | 439 | 106 | 1,068 |
| 100 | 60 | 39 | 23 | 156 | 154 | 129 | 404 | 104 | 1,009 |
| 100 | 70 | 39 | 23 | 155 | 144 | 128 | 352 | 97 | 938 |
| 100 | 80 | 39 | 23 | 149 | 129 | 126 | 307 | 95 | 868 |
| 100 | 90 | 39 | 23 | 133 | 110 | 121 | 228 | 83 | 737 |
| 100 | 100 | 39 | 23 | 114 | 109 | 120 | 178 | 80 | 663 |

The overlapping threshold for merging and splitting events was set to 50%. The total number of events found by Asur et al. method was 747, out of which 120 were continuation, 23 were forming, 16 were dissolving, 255 were merging and 333 were splitting.

Again, the small number of continuing, forming, and dissolving events is caused by the too rigorous conditions. In turn, the great number of merging (splitting) events is a result of low overlapping threshold for merge (split).

As in case of CPM grouping method, the number of events found on data grouped by the *Blondel* method is also overestimated. The number of anomalies this time is 40 cases, 5% of all results. Therefore, the total number of "distinct" events was 707. This may mean that Asur et al. method works slightly better for disjoint groups.

### 5.2.2 The GED method

As previously, for the data grouped with the CPM method, the GED method have been run with different values of $\alpha$ and $\beta$ thresholds and the results are presented in Table 3. The time needed for a single run was about 13 min. The thresholds for the *forming* and *dissolving* event was again set to 10%.

The total number of events found with thresholds equal to 50% was 1,231 but with thresholds equal to 100% only 663. This indicates that parameters α and β influence the number of events even more than in case of the CPM method. The linear relation between the increase of threshold α or β and the reduction of the number of merging (splitting) is preserved (Table 3; Figs. 11, 12, 13, 14, 15, 16).

The GED method found 120 continue events when both inclusions of groups (α and β) are equal to 100%, which correspond to *continuation* event in Asur et al. method.

In general, the GED method can be successfully used for both, overlapping or disjoint groups. If overlapping groups for a small network are needed then CPM can be used, however, if one needs to extract groups very fast and for a big network then the method proposed by *Blondel* can be utilized. This flexibility and adaptability of the GED method is its big advantage because most methods can be used only for either overlapping or disjoint groups.

### 5.2.3 Differences between the GED method and the method by Asur et al.

The GED method needed less than 8 h to calculate results for the whole Table 3, while a single run of Asur et al. method lasted almost 6 h. A single run of the GED method was only 13 min, so it is still much faster than the method by Asur et al.

The GED method run with thresholds equals 50% found as many as 613 events, which the method by Asur et al. did not recognize at all. Again, the big gap in results obtained with Asur et al. method is caused mostly by its rigorous conditions for assigning events and almost no flexibility of the method. Like in case of the CPM method, Asur et al. method found events, which the GED method skipped because of threshold values. Again, these events were found after reducing the thresholds.

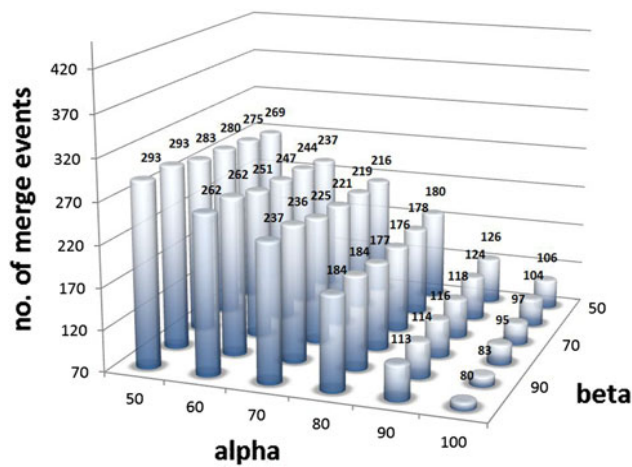

Fig. 12 Alpha and beta influence on the number of merge events

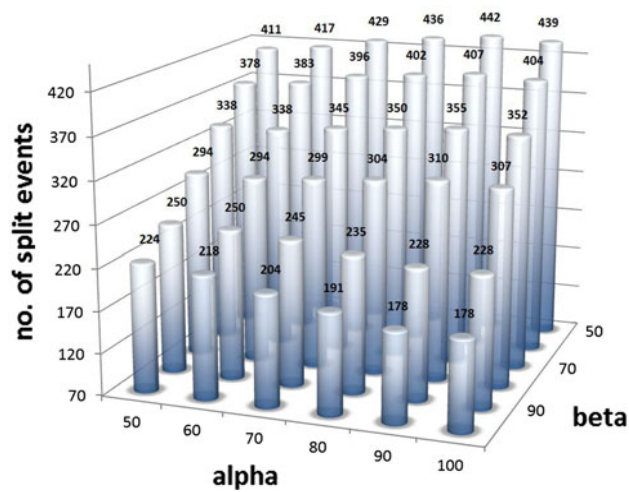

Fig. 13 Alpha and beta influence on the number of split events

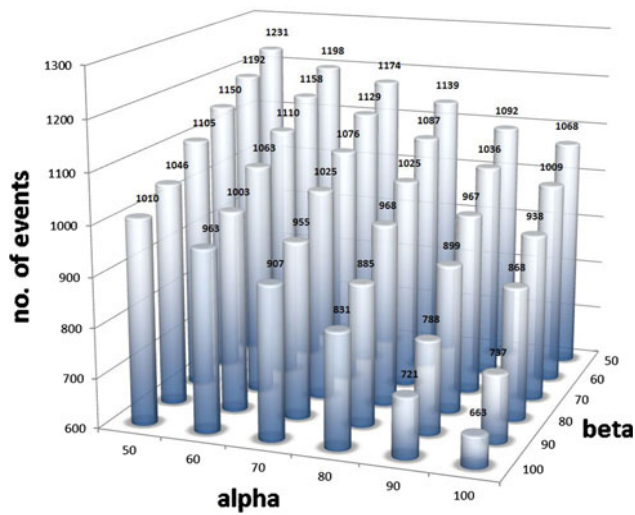

Fig. 11 Alpha and beta influence on the number of events

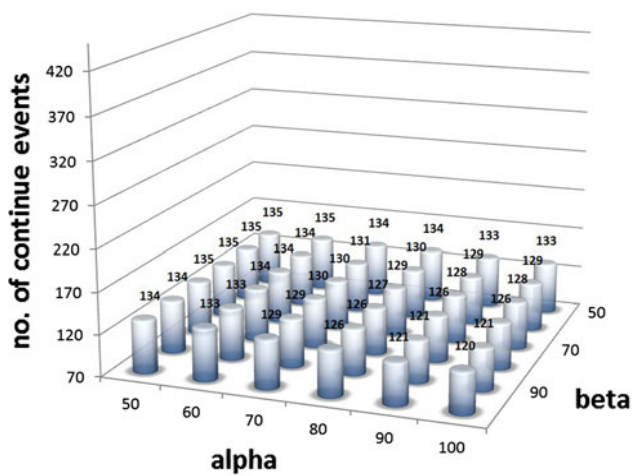

Fig. 14 Alpha and beta influence on the number of continue events

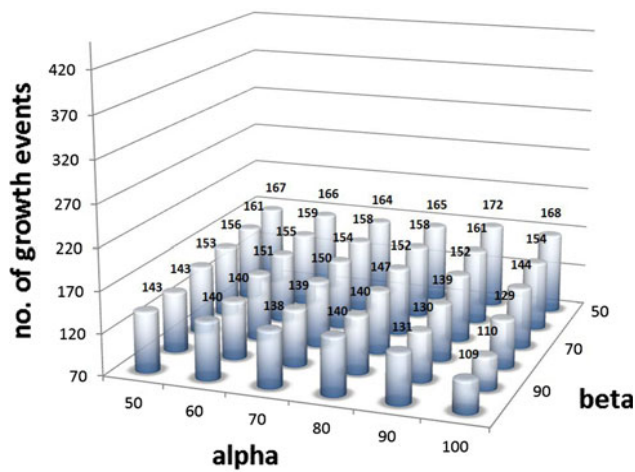

**Fig. 15** Alpha and beta influence on the number of growth events

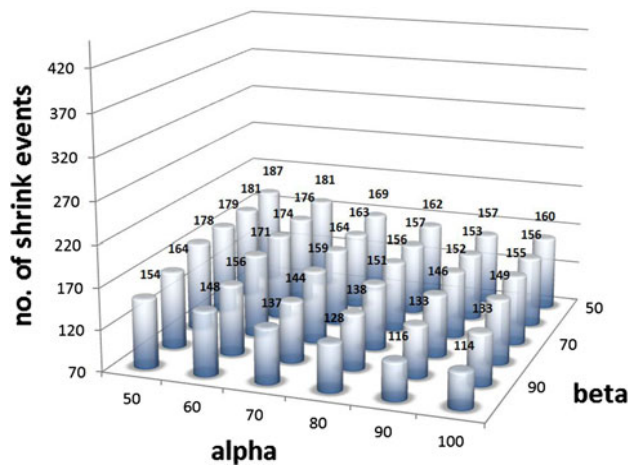

**Fig. 16** Alpha and beta influence on the number of shrink events

The above considerations confirm that the GED method is better than Asur et al. method for overlapping as well as for disjoint methods of grouping.

## 6 Conclusions

The increasing number of systems in which people communicate with each other continues to rise. That creates an insatiable need and opportunity to analyse them. An important part of such studies is group extraction and analysis of their evolution over time which help in understanding the mechanisms governing the development and variability of social groups.

The GED method, proposed in the paper, uses not only the size and equivalence of groups' members, but it also takes into account their position and importance within the group in order to identify what happened with the group in the successive time frames. This was mainly achieved by the new measure called *inclusion*, which respects both the quantity (the number of members) and quality (the importance of members) of the group. The GED method was designed to be as much flexible as possible and to fit in to both overlapping and non-overlapping groups. Simultaneously, it preserves the low and adjustable computational complexity. Because of many different user importance (centrality) indicators, which can be used in the *inclusion* measure and owing to its two parameters $\alpha$ and $\beta$, full control over the method is provided.

The results of experiments and comparison with two existing methods presented in Sect. 5, leads to the conclusion that the desired effects were achieved, and the new GED method may become one of the best methods for group evolution discovery.

**Acknowledgments** The work was partially supported by The Polish Ministry of Science and Higher Education the research project 2010–2013 and Fellowship co-Financed by European Union within European Social Fund. The authors also would like to thank Mr Gergely Palla for his help in understanding his method.

## References

Agarwal N, Galan M, Liu H, Subramanya S (2010) WisColl: collective Wisdom based Blog Clustering. Inf Sci 180(1):39–61

Amaral LAN, Scala A, Barthelemy M, Stanley HE (2000) Classes of small-world networks. Proc Natl Acad Sci USA 97:11149–11152

Asur S, Parthasarathy S, Ucar D (2007) An event-based framework for characterizing the evolutionary behavior of interaction graphs. In: Proceedings of the 13th ACM SIGKDD international conference on knowledge discovery and data mining (KDD '07). ACM, New York, NY, USA, pp 913–921

Barrat A, Barthelemy M, Vespignani A (2008) Dynamical processes on complex networks. Cambridge University Press, UK

Blondel VD, Guillaume JL, Lambiotte R, Lefebvre E (2008) Fast unfolding of communities in large network. J Stat Mech Theory Exp P10008

Bródka P, Musial K, Kazienko P (2009) A performance of centrality calculation in social networks. In: Proceedings of the 2009 international conference on computational aspects of social networks (CASON '09). IEEE Computer Society, Washington, DC, USA, pp 24–31

Bródka P, Saganowski S, Kazienko P (2011) Group evolution discovery in social networks. In: ASONAM 2011, the 2011

international conference on advances in social network analysis and mining, Kaohsiung, Taiwan, 25–27 July 2011, IEEE Computer Society, pp 247–253

Bródka P, Saganowski P, Kazienko P (2011) Tracking group evolution in social networks. SocInfo '11, The third international conference on social informatics, 6–8 October 2011, Singapore, Lecture Notes in Artificial Intelligence LNAI, Springer, pp 316–319

Cattell V (2001) Poor people, poor places, and poor health: the mediating role of social networks and social capital. Soc Sci Med 52(10):1501–1516

Chakrabarti D, Kumar R, Tomkins A (2006) Evolutionary clustering. In: Proceedings of the 12th ACM SIGKDD international conference on knowledge discovery and data mining KDD '06, Philadelphia, Pennsylvania, USA

Cheng X, Dale C, Liu J (2008) Statistics and social networking of 'YouTube videos'. In: Proceedings of the 16th international workshop on quality of service, IEEE, pp 229–238

Coleman JS (1964) An introduction to mathematical sociology. Collier-Macmillan, London

Derényi I, Palla G, Vicsek T (2005) Clique percolation in random networks. Phys Rev Lett 94:160202

Ellison NB, Steinfield C, Lampe C (2007) The benefits of Facebook "friends": social capital and college students' use of online social network sites. J Comput Mediat Commun 12(4):article 1. http://jcmc.indiana.edu/vol12/issue4/ellison.html

Fortunato S (2010) Community detection in graphs. Phys Rep 486(3–5):75–174

Freeman LC (2004) The development of social network analysis: a study in the sociology of science. BookSurge Publishing

Garton L, Haythorntwaite C, Wellman B (1997) Studying online social networks. J Comput Mediat Commun 3(1):75–105. http://jcmc.indiana.edu/vol3/issue1/garton.html

Girvan M, Newman MEJ (2002) Community structure in social and biological networks. Proc Natl Acad Sci USA 99(12):7821–7826

Golbeck J, Hendler J (2006) FilmTrust: movie recommendations using trust in web-based social networks. In: Proceedings of consumer communications and networking conference, IEEE conference proceedings, vol 1, pp 282–286

Hanneman R, Riddle M (2005) Introduction to social network methods, online textbook. University of California, Riverside, CA. http://faculty.ucr.edu/~hanneman/nettext/

Hopcroft J, Khan O, Kulis B, Selman B (2004) Tracking evolving communities in large linked networks. Proc Natl Acad Sci PNAS USA 101:5249

Huberman B, Romero D, Wu F (2009) Social networks that matter: Twitter under the microscope. First Monday, pp 1–5 (arXiv:0812.1045v1)

Kazienko P, Ruta D, Bródka P (2009a) The impact of customer churn on social value dynamics. Int J Virtual Communities Soc Netw 1(3):60–72

Kazienko P, Musiał K, Zgrzywa A (2009b) Evaluation of node position based on email communication. Control Cybern 38(1):67–86

Kazienko P, Musial K, Kajdanowicz T (2011) Multidimensional social network and its application to the social recommender system. IEEE Trans Syst Man and Cybern Part A Syst Hum 41(4):746–759

Kim MS, Han J (2009) A particle and density based evolutionary clustering method for dynamic networks. In: Proceedings of 2009 international conferance on very large data bases, Lyon, France

Kottak CP (2004) Mirror for humanity: a concise introduction to cultural anthropology, McGraw-Hill, New York, USA

Lazega E (2001) The collegial phenomenon. The social mechanism of co-operation among peers in a corporate law partnership. Oxford University Press, Oxford

Lin YR, Chi Y, Zhu S, Sundaram H, Tseng BL (2008) Facetnet: a framework for analyzing communities and their evolutions in dynamic networks. In: Proceeding of the 17th international conference on World Wide Web, April 21–25, Beijing, China

Montgomery J (1991) Social networks and labor-market outcomes: toward an economic analysis, American Economic Review 81, vol 5, pp 1407–1418

Moody J, White DR (2003) Structural cohesion and embeddedness: a hierarchical concept of social groups. Am Sociol Rev 68(1):103–127

Morris M (1997) Sexual network and HIV. AIDS 11:209–216

Musial K, Kazienko P, Bródka P (2009) User position measures in social networks. In: Proceedings of the 3rd workshop on social network mining and analysis (SNA-KDD '09). ACM, New York, NY, USA, Article 6

Newman MEJ (2001) The structure of scientific collaboration networks. Proc Natl Acad Sci USA 98:404–409

Pagel M, Erdly W, Becker J (1987) Social networks: we get by with (and in spite of) a little help from our friends. J Pers Soc Psychol 53(4):793–804

Palla G, Derényi I, Farkas I, Vicsek T (2005) Uncovering the overlapping community structure of complex networks in nature and society. Nature 435:814–818

Palla G, Barabási AL, Vicsek T (2007) Quantifying social group evolution. Nature 446:664–667

Robins GL, Alexander M (2004) Small worlds among interlocking directors: network structure and distance in bipartite graphs. Comput Math Organ Theory 10(1):69–94

Scott J (2000) Social network analysis: a handbook. SAGE, London

Sun J, Papadimitriou S, Yu PS, Faloutsos C, GraphScope (2007) Parameter-free mining of large time-evolving graphs. In: Proceedings of the 13th ACM SIGKDD international conference on knowledge discovery and data mining (KDD '07). ACM, New York, NY, USA, pp 687–696

Traud AL, Kelsic ED, Mucha PJ, Porter MA (2009) Community structure in online collegiate social networks. eprint arXiv:0809.0690

Tyler JR, Wilkinson DM, Huberman BA (2003) Email as spectroscopy: automated discovery of community structure within organizations. In: Communities and technologies. Kluwer, Deventer, pp 81–96

Wasserman S, Faust K (1994) Social network analysis: methods and applications. Cambridge University Press, New York

Watts DJ, Strogatz S (1998) Collective dynamics of 'small-world' networks. Nature 393:440–444

Wellman B, Salaff J, Dimitrova D, Garton L, Gulia M, Haythornthwaite C (1996) Computer networks as social networks: collaborative work telework, and virtual community. Annu Rev Sociol 22(1):213–238